\documentclass[aps,prl,showpacs,twocolumn,amsmath,amssymb,amsfonts,superscriptaddress]{revtex4}

\usepackage{epsfig,graphicx,bm}

\newcommand{\beq}{\begin{equation}}
\newcommand{\eeq}{\end{equation}}
\newcommand{\beqa}{\begin{eqnarray}}
\newcommand{\eeqa}{\end{eqnarray}}
\newcommand{\ket} [1] {\vert #1 \rangle}
\newcommand{\bra} [1] {\langle #1 \vert}

\newcommand{\proj}[1]{\ket{#1}\bra{#1}}
\newcommand{\mean}[1]{\langle #1 \rangle}

\begin{document}

\title{Quantum entanglement enhances the capacity of bosonic channels with memory}
\author{Nicolas J. Cerf}
\author{Julien Clavareau}
\affiliation{QUIC, Ecole Polytechnique, CP 165/59,
Universit\'e Libre de Bruxelles, B-1050 Brussels, Belgium}
\author{Chiara Macchiavello}
\affiliation{QUIT, Dipartimento di Fisica ``A. Volta'', Universit\`a di Pavia,
via Bassi 6, I-27100 Pavia, Italy}
\author{J\'er\'emie Roland}
\affiliation{QUIC, Ecole Polytechnique, CP 165/59,
Universit\'e Libre de Bruxelles, B-1050 Brussels, Belgium}
\affiliation{Laboratoire de Recherche en Informatique, UMR 8263, Universit\'e de Paris-Sud, 91405 Orsay, France}

\begin{abstract}
The bosonic quantum channels have recently attracted a growing interest, 
motivated by the hope that they open a tractable approach 
to the generally hard problem of evaluating quantum channel capacities.
These studies, however, have always been restricted to memoryless channels. 
Here, it is shown that the classical capacity of a bosonic Gaussian channel
with memory can be significantly enhanced if entangled symbols are
used instead of product symbols. For example,
the capacity of a photonic channel with 70\%-correlated thermal noise
of one third the shot noise
is enhanced by about 11\% when using 3.8-dB entangled light with a modulation
variance equal to the shot noise.
\end{abstract}


\maketitle

A main goal of quantum information theory is to evaluate the information capacities of quantum communication channels. In particular, an important
question is to determine how much {\em classical} information can be processed
asymptotically via a quantum channel. This problem has been solved, today,
only for a few quantum channels, and it has been addressed only recently
for bosonic channels, i.e., continuous-variable quantum channels acting
on a bosonic field such as the electromagnetic field \cite{holevo-werner}.
The classical capacity of a purely {\em lossy} bosonic channel was solved exactly very recently \cite{lossy-capacity}, while  
the case of {\em noisy} bosonic channels is already more involved. Actually,
the classical capacity of the Gaussian bosonic channel, i.e., 
a continuous-variable quantum channel undergoing a Gaussian-distributed 
thermal noise, has been derived in \cite{holevoetal}
although this result only holds provided that the optimal input ensemble 
is a tensor product of Gaussian states, as conjectured by several authors
but not rigorously proven today (see e.g. \cite{giovannetti} for recent progress
on this problem). All these studies, however, have been restricted to memoryless bosonic channels.
\par

In this Letter, we investigate the capacity of a bosonic Gaussian channel 
that exhibits memory. This study is motivated by the recent finding that,
for some appropriate extension of the depolarizing channel with correlated noise, entangled qubit pairs can enhance the 2-shot classical capacity
\cite{macch-palma}. Here, we consider channels with a thermal noise
that has a {\em finite} bandwidth.
The resulting memory effect is modeled by assuming that the noise
affecting two subsequent uses of the channel follows a bivariate Gaussian
distribution with a non-vanishing correlation coefficient, measuring the
degree of memory of the channel. We prove that if the memory
is non-zero and if the input energy is constrained, 
then the channel capacity can be significantly
enhanced by using entangled symbols instead of product symbols,
in contrast with the common knowledge that entanglement is of no use
for information transfer via a quantum channel. 
The relation between the degree of memory and the resulting optimal input entanglement is analyzed.
\par

\paragraph{Bosonic Gaussian channels.}
Let us define a memoryless bosonic Gaussian channel $T$ 
acting on a mode of the electromagnetic field associated with
the annihilation and creation operators $a$ and $a^\dagger$,
or, equivalently, the quadrature components $q=(a+a^\dagger)/\sqrt{2}$
and $p=i(a^\dagger-a)/\sqrt{2}$, satisfying
the commutation relation $[q,p]=i$.
If the input of the channel is initially in state $\rho$, we have
\beq
\rho\mapsto T[\rho]=\int d^2\beta\ q(\beta)\ D(\beta)\rho D^\dagger(\beta),
\eeq
where $d^2\beta=d\Re(\beta) \, d\Im(\beta)$, 
while $D(\beta)=e^{\beta a^\dagger-\beta^* a}$ denotes 
the displacement operator (such that $\ket{\alpha}=D(\alpha)\ket{0}$
with $\ket{0}$ being the vacuum state and $\ket{\alpha}$ being a coherent 
state of mean value $\alpha$). For a Gaussian channel, the kernel
is a bivariate Gaussian distribution with variance $N$, that is,
$q(\beta)=\frac{1}{\pi N} \, e^{-\frac{|\beta|^2}{N}}$.
The channel then randomly displaces an input coherent state
according to a Gaussian distribution, which results 
in a thermal state ($N$ is the variance of the added noise
on the quadrature components $q$ and $p$, or, equivalently
the number of thermal photons added by the channel). 
The Gaussian CP map effected by this channel
can also be characterized via the covariance matrix.
Restricting to Gaussian states with a vanishing mean value, 
a complete state characterization is provided by the covariance matrix
\beqa
\gamma&=&\left(\begin{array}{cc}
\mean{q^2} & \frac{1}{2}\mean{qp+pq} \\ 
\frac{1}{2}\mean{qp+pq} & \mean{p^2}
\end{array} \right).
\eeqa
The Gaussian channel can then be written as
\beq
\gamma \mapsto \gamma + \left(\begin{array}{cc}
N & 0 \\ 
0 & N
\end{array} \right).
\eeq

\paragraph{Classical capacity of a quantum channel.}
The coding theorem for quantum channels asserts 
that the one-shot classical capacity of a quantum
channel $T$ is given by
\beq\label{capacity}
C_1(T)=\max\left[S\left(\sum_i p_i T[\rho_i]\right)-\sum_i p_i S\left(T[\rho_i]\right)\right],
\eeq
where 
$S(\rho)=-\text{Tr}(\rho\log\rho)$
is the von Neumann entropy of the density operator $\rho$.
In Eq.~(\ref{capacity}),
the maximum is taken over all probability distributions $\{p_i\}$ and collections of density operators $\{\rho_i\}$
satisfying the energy constraint
\beq
\sum_i p_i \text{Tr}\left(\rho_i a^\dagger a\right)\leq\bar{n},
\eeq
with $\bar{n}$ being the maximum mean photon number at the input
of the channel. For a monomodal bosonic Gaussian channel,
it is conjectured that a Gaussian mixture of coherent states
(i.e., a thermal state) achieves the channel capacity \cite{fn1}.
The sum over $i$ is replaced by an integral over $\alpha$,
where the input states $\rho_\alpha^{\text{in}}=\proj{\alpha}$ are
drawn from the probability density
$p(\alpha)=\frac{1}{\pi \bar{n}} \, e^{-\frac{|\alpha|^2}{\bar{n}}}$.
Thus, the one-shot classical capacity of the channel becomes
\beq
C_1(T)=S\left(\bar{\rho}\right)-\int d^2\alpha \,  p(\alpha) \, S\left(\rho_\alpha^{\text{out}}\right),
\eeq
where we have defined the individual output states
\beq
\rho_\alpha^{\text{out}}=T[\rho_\alpha^{\text{in}}]=\frac{1}{\pi N}\int d^2\beta \, e^{-\frac{|\beta-\alpha|^2}{N}} \, \proj{\beta}
\eeq
and their mixture (saturating the energy constraint)
\beq
\hspace*{-2mm}\bar{\rho}=\int d^2\alpha \, p(\alpha) \, \rho_\alpha^{\text{out}}
=\frac{1}{\pi(\bar{n}+N)}\int d^2\beta \, e^{-\frac{|\beta|^2}{\bar{n}+N}} \, \proj{\beta}.
\eeq
In order to calculate the entropy of a state $\rho$, 
one computes the symplectic values of its covariance matrix
$\gamma$, i.e., the solutions 
of the equation $\left|\gamma-\lambda J\right|=0$,
where
\beq\label{matrix-j}
J=\left(\begin{array}{cc}
0 & i \\ 
-i & 0
\end{array} \right).
\eeq
It can be shown that these values always come as a pair $\pm\lambda$, so that
the entropy is given by $S(\rho)=g\left(|\lambda|-\frac{1}{2}\right)$,
where
\beq
g(x)= \left\{ \begin{array}{ll}
(x+1)\log_2 (x+1)-x\log_2 x,\qquad & x>0\\
0\qquad & x=0 \end{array} \right.
\eeq
is the entropy of a thermal state with a mean photon number of $x$.
Since the input states $\rho_\alpha^{\text{in}}$ are coherent states with a covariance matrix
\beq\label{matrix-coherent}
\gamma^\text{in}=\frac{1}{2}
\left(\begin{array}{cc}
1 & 0 \\ 
0 & 1
\end{array} \right),
\eeq
the individual output states $\rho_\alpha^\text{out}$ and their mixture $\bar{\rho}$ are associated with the covariance matrices
\beqa
\gamma^\text{out}&=&\frac{1}{2}
\left(\begin{array}{cc}
1+2N & 0 \\ 
0 & 1+2N
\end{array} \right),\\
\bar{\gamma}&=&\frac{1}{2}
\left(\begin{array}{cc}
1+2(\bar{n}+N) & 0 \\ 
0 & 1+2(\bar{n}+N)
\end{array} \right),
\eeqa
so that the one-shot capacity of the channel is
\beq
C_1(T)=g(\bar{n}+N)-g(N).
\eeq

\paragraph{Bimodal channel.}
Consider two subsequent uses of a memoryless channel $T$, 
defining the bimodal channel 
\beqa
\lefteqn{\rho \mapsto T_{12}[\rho]=\int d^2\beta_1 \ d^2\beta_2 \  q(\beta_1,\beta_2)} \hspace{0.5cm} \nonumber \\
&\times& D(\beta_1)\otimes D(\beta_2)\; \rho \; D^\dagger(\beta_1)\otimes D^\dagger(\beta_2),
\eeqa
where
$q(\beta_1,\beta_2)=\frac{1}{\pi^2 N^2} e^{-\frac{|\beta_1|^2+|\beta_2|^2}{N}}$
since the noise affecting the two uses is uncorrelated.
Ordering the quadrature components of the two modes in a column vector
$R=[q_1,p_1,q_2,p_2]^\text{T}$, we 
define the covariance matrix $\gamma_{12}$ of a bimodal state $\rho_{12}$ as
\beq
\gamma_{12}=\text{Tr}(R\, \rho_{12} \, R^\text{T})- \textstyle{\frac{1}{2}}
 \, J_1 \oplus J_2,
\eeq
where each $J_j$ takes the form (\ref{matrix-j}).
We restrict ourselves to bimodal Gaussian states \cite{fn1},
characterized by 
\beq\label{generic-covariance}
\gamma_{12}=\left(\begin{array}{cc}
\gamma_1 & \sigma_{12} \\ 
\sigma_{12}^\text{T} & \gamma_2
\end{array}\right),
\eeq
where $\gamma_1$ is the covariance matrix associated
with the reduced density operator $\rho_1=\text{Tr}_2(\rho_{12})$ 
of mode $1$ (and similarly for $\gamma_2$),
while $\sigma_{12}$ characterizes the correlation and/or entanglement 
between the two modes. For a memoryless channel,
the optimal input states are simply products of coherent states, 
with a covariance matrix
$\gamma_{12}^\text{in}=\gamma_1^\text{in}\oplus\gamma_2^\text{in}$
where $\gamma_1^\text{in}$ and $\gamma_2^\text{in}$ both take the form (\ref{matrix-coherent}), while $\sigma_{12}^\text{in}=0$. 
The optimal input modulation is a product of Gaussian distributions, $p(\alpha_1,\alpha_2) = \frac{1}{\pi^2 \bar{n}^2} e^{-\frac{|\alpha_1|^2+|\alpha_2|^2}{\bar{n}}}$.
It follows that the classical capacity of this channel is additive
$\frac{1}{2} \times C_1(T_{12})=C_1(T)$.

\paragraph{Bosonic Gaussian channel with memory.}
Let us investigate what happens
if the noise is correlated, for instance when the two uses 
are closely separated in time and the channel has a finite bandwidth.
We assume that the noise distribution
takes the general form
\beq
q(\beta_1,\beta_2)=\frac{1}{\pi^2 \sqrt{|\gamma_N|}}
e^{-\bm{\beta}^\dagger\gamma_N^{-1}\bm{\beta}},
\eeq
where 
$\bm{\beta}=[\Re(\beta_1),\Im(\beta_1),\Re(\beta_2),\Im(\beta_2)]^\text{T}$
and $\gamma_N$ is the covariance matrix of the noise quadratures,
chosen to be
\beq
\gamma_N=\left(\begin{array}{cccc}
N & 0 & -xN & 0 \\ 
0 & N & 0 & xN \\ 
-xN & 0 & N & 0 \\ 
0 & xN & 0 & N
\end{array} \right).
\eeq
Thus, the map $T_{12}$ can be expressed by
$\gamma_{12} \mapsto \gamma_{12} + \gamma_N$,
so that the noise terms added on the $p$ quadratures 
of modes 1 and 2 are correlated Gaussians with variance $N$ (those added
on the $q$ quadratures are anticorrelated Gaussians with variance $N$
\cite{fn0}).
The correlation coefficient $x$ ranges from
$x=0$ for a memoryless channel to $x=1$ for a channel
with full memory.
\par

We now come to the central result of this paper. While we have seen that
for a memoryless channel, the capacity is
attained for product states, we will prove that for correlated thermal noise,
the capacity is achieved if some appropriate degree of entanglement
is injected at the input of the channel. 
Intuitively, if we take an EPR state, i.e., the common eigenstate
of $q_1 +q_2$ and $p_1-p_2$ with respective eigenvalues $q_+$ and $p_-$,
it is clear that the noise on $q_+$ and $p_-$ effected by the channel
is reduced as $x$ increases. This suggests that using entangled
input states may decrease the effective noise, hence increase the capacity.
However, EPR states have infinite energy so they violate
the energy constraint.
Instead, we may inject (finite-energy) two-mode vacuum squeezed states,
whose covariance matrix is given by
\beqa
\gamma_1^\text{in}&=&\gamma_2^\text{in}=\frac{1}{2}\left(
\begin{array}{cc}
\cosh 2r & 0 \\ 
0 & \cosh 2r
\end{array} 
\right),   \label{EPR1}\\
\sigma_{12}^\text{in}&=&\frac{1}{2}\left(
\begin{array}{cc}
-\sinh 2r & 0 \\ 
0 & \sinh 2r
\end{array} 
\right),  \label{EPR2}
\eeqa
with $r$ being the squeezing parameter.
Note that purely classical correlations between 
the quadratures in the input distribution $p(\alpha_1,\alpha_2)$
also help increase the capacity when $x>0$, so we have to check
that entanglement gives an extra enhancement in addition to this.
\par

The mean photon number in each mode of the state characterized by 
Eqs.~(\ref{EPR1})-(\ref{EPR2}) is $\sinh^2 r$, 
so that the maximum allowed modulation (for a fixed maximum photon number
$\bar{n}$) decreases as entanglement increases. Remarkably, there 
is a possible compromise between this reduction of modulation 
and the entanglement-induced noise reduction on $q_+$ and $p_-$. 
To show this, consider input states with $\sinh^2 r = \eta \bar{n}$,
where $\eta$ measures the {\em degree of entanglement} 
and is used to interpolate between
a product of vacuum states ($\eta=0$), which can be maximally modulated,
and an entangled state ($\eta=1$), for which the entire
energy is due to entanglement and no modulation can be applied.
At the output of the channel, we get 
states with a covariance matrix $\gamma_{12}^\text{out}$ where
\beqa
\gamma_{1,2}^\text{out}&=&\frac{1}{2}\left(
\begin{array}{cc}
\cosh 2r+2N & 0 \\ 
0 & \cosh 2r+2N
\end{array} 
\right),   \label{gammaout}\\
\sigma_{12}^\text{out}&=&\frac{1}{2}\left(
\begin{array}{cc}
-\sinh 2r-2xN & 0 \\ 
0 & \sinh 2r+2xN
\end{array} 
\right),
\eeqa
while the mixture $\bar{\gamma}_{12}$ of these states are characterized by
\beqa
\bar{\gamma}_{1,2}&=&
\gamma_{1,2}^\text{out}+\left(
\begin{array}{cc}
(1-\eta)\bar{n} & 0 \\ 
0 & (1-\eta)\bar{n}
\end{array} 
\right),\\
\bar{\sigma}_{12}&=& \sigma_{12}^\text{out} +\left(
\begin{array}{cc}
y (1-\eta)\bar{n} & 0 \\ 
0 & -y (1-\eta)\bar{n}
\end{array} 
\right) ,   \label{sigmamean}
\eeqa
assuming that the energy constraint is saturated. Here,
$y$ stands for the classical input correlation coefficient
(to compensate for the noise,
the $q$ displacements need to be correlated, 
and the $p$ displacements anti-correlated).

\begin{figure}[t]
\begin{center}
\epsfig{figure=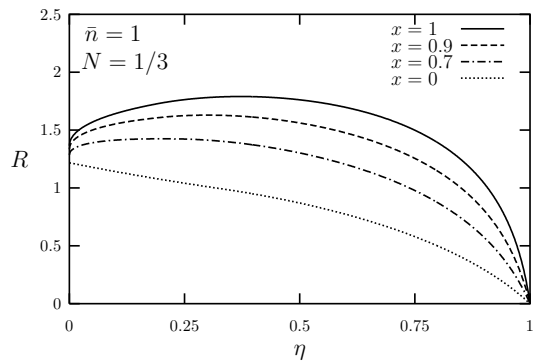,width=70mm}
\caption{Transmission rate $R$ (maximized over the classical
correlation coefficient $y$)
as a function of the input entanglement $\eta$
for a thermal channel with a degree of memory $x$. The mean
number of photons is $\bar{n}=1$ at the input, while the number of added
thermal photons is $N=1/3$.}
\label{fig1}
\end{center}
\end{figure}

\paragraph{Entangled-enhanced capacity.}
In order to evaluate the transmission rate achieved by these states, we need
first to compute the symplectic values $\lambda_{12}^\text{out}$
and $\bar{\lambda}_{12}$ of $\gamma_{12}^\text{out}$
and $\bar{\gamma}_{12}$, respectively. 
The symplectic values $\pm\lambda_{12}$ of a covariance matrix $\gamma_{12}$ of the generic form
(\ref{generic-covariance}) are the solutions of the equation
$\left|\gamma_{12}-\lambda_{12} (J_1\oplus J_2)\right|=0$,
or, equivalently, the biquadratic equation
\beq
\lambda_{12}^4-(|\gamma_1|+|\gamma_2|+2|\sigma_{12}|)\lambda_{12}^2
+|\gamma_{12}|=0.
\eeq
Using Eqs.~(\ref{gammaout})-(\ref{sigmamean}), we see that
$\gamma_{12}^\text{out}$ and $\bar{\gamma}_{12}$ admit each one pair of doubly-degenerate symplectic values, namely
$\lambda_{12}^\text{out}=\pm \sqrt{u_\text{out}^2-v_\text{out}^2}$
and $\bar{\lambda}_{12}=\pm \sqrt{\bar{u}^2-\bar{v}^2}$, with
\beqa
u_\text{out}= \textstyle{\frac{1}{2}}+\eta\bar{n}+N, \hspace{1.2cm}
v_\text{out}=\sqrt{\eta\bar{n}(1+\eta\bar{n})}+xN , \nonumber\\
\bar{u}=\textstyle{\frac{1}{2}}+\bar{n}+N, \quad
\bar{v}=\sqrt{\eta\bar{n}(1+\eta\bar{n})}+xN-y(1-\eta)\bar{n}.\nonumber
\eeqa
The transmission rate per mode is then given by
\beq
R(y,\eta)
=g(|\bar{\lambda}_{12}|-\textstyle{\frac{1}{2}})-g(|\lambda_{12}^\text{out}|-\textstyle{\frac{1}{2}})
\eeq
When $x > 0$, the optimized rate $R$ over $y$ increases with the degree of entanglement $\eta$
and attains a maximum at some optimal value $\eta^* > 0$ (see Fig.~\ref{fig1}),
so that the maximum is achieved by entangled input states as advertized.
\par

\begin{figure}[t]
\begin{center}
\epsfig{figure=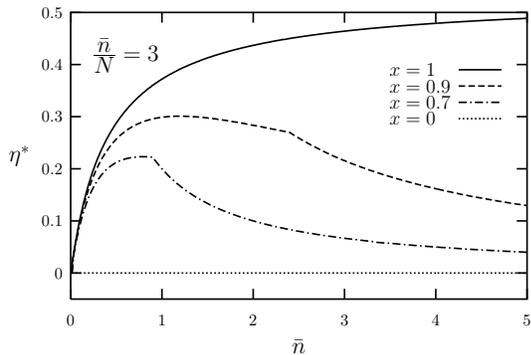,width=70mm}
\caption{Optimal degree of input entanglement $\eta^*$ as a function
of the mean input photon number $\bar{n}$ for a fixed signal-to-noise
ratio $\bar{n}/N=3$.}
\label{fig2}
\end{center}
\end{figure}

\begin{figure}[t]
\begin{center}
\epsfig{figure=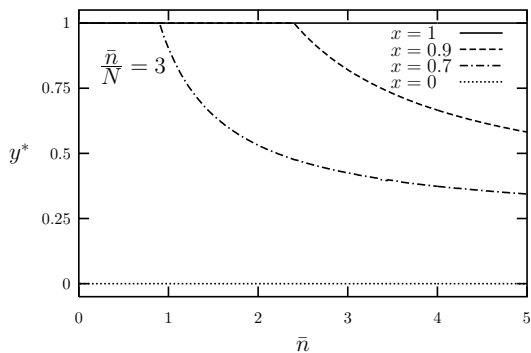,width=70mm}
\caption{Optimal degree of input correlation $y^*$ as a function
of the mean input photon number $\bar{n}$ for a fixed signal-to-noise ratio $\bar{n}/N=3$.}
\label{fig3}
\end{center}
\end{figure}

It now suffices to maximize $R$ with respect to both $y$ and $\eta$ 
in order to find the channel capacity $C$ (assuming that the conjecture \cite{fn1} is verified and that no product but non-Gaussian states may outperform the Gaussian entangled states considered here).
If we keep the signal-to-noise
ratio $\bar{n}/N$ constant, it is visible from Fig.~\ref{fig2} that
the optimal degree of entanglement $\eta^*$ is the highest 
at some particular value of the mean input photon number $\bar{n}$, 
and then decreases back to zero
in the large-$\bar{n}$ limit (except if $x=0$ or $1$).
Clearly, in this limit, the channel $T$ tends to a couple of 
classical channels with Gaussian additive noise (one for each quadrature),
so that entanglement cannot play a role any more \cite{fn2}. 
Fig.~\ref{fig3} shows the corresponding optimal value
of the input correlation coefficient $y^*$ 
for the same values of the other relevant parameters. Note that,
even in the classical limit $\bar{n}\to\infty$,
some non-zero input correlation is useful to enhance the capacity 
of a Gaussian channel with $x>0$.

\begin{figure}[h]
\begin{center}
\epsfig{figure=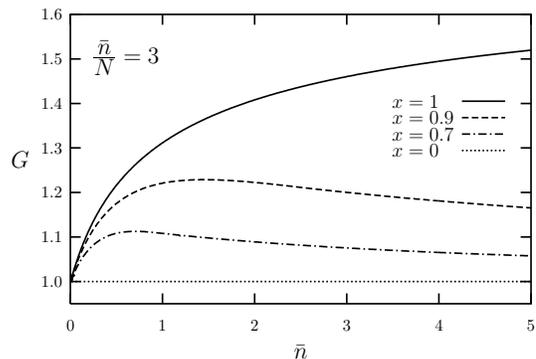,width=70mm}
\caption{Capacity gain $G=\max_{y,\eta}R(y,\eta)/\max_y R(y,0)$
as a function of the mean input photon number $\bar{n}$
for a fixed signal-to-noise ratio $\bar{n}/N=3$.}
\label{fig4}
\end{center}
\end{figure}

\paragraph{Conclusions.}
We have shown that entangled states can be used to enhance
the classical capacity of a bosonic channel undergoing a thermal noise with
memory. We determined the amount of entanglement that maximizes 
the information transmitted over the channel for a given input energy constraint
(mean photon number per mode) and a given noise level (mean number of thermal photons per mode). For example, the capacity of a channel with a mean number 
of thermal photons of 1/3 and a correlation coefficient of 70\% is enhanced
by 10.8\% if the mean photon number is 1 and the two-mode squeezing is 3.8 dB
at the input. This capacity enhancement may seem paradoxical 
at first sight since using entangled signal states necessarily
decreases the modulation variance for a fixed input energy, which
seemingly lowers the capacity. However, due to the quantum correlations
of entangled states, the noise affecting one mode can be partly
compensated by the correlated noise affecting the second mode, 
which globally reduces the effective noise. 
Interestingly, there exists a regime in which this latter effect dominates,
resulting in a net enhancement of the amount of classical information
transmitted per use of the channel.
The capacity gain $G$, measuring the entanglement-induced capacity enhancement, 
is plotted in Fig.~\ref{fig4}. It illustrates that a capacity enhancement 
of tens of percents is achievable by using entangled light beams with
experimentally accessible levels of squeezing.

We thank V. Giovannetti for informing us of his recent
related work on bosonic memory channels, see e-print quant-ph/0410176.
We are also very grateful to A. Holevo for useful comments.
We acknowledge financial support from the Communaut\'e Fran\c caise 
de Belgique under grant ARC 00/05-251, from the IUAP programme of the Belgian government under grant V-18, and from the EU under project COVAQIAL.
J.R. acknowledges support from the Belgian FRIA foundation.

\begin{thebibliography}{99}
\bibitem{fn1} In this work, we do not question the generally admitted
conjecture that Gaussian states achieve the classical capacity, so we
restrict our analysis to Gaussian input states.
\bibitem{fn0} The beneficial effect of entanglement disappears
in a more symmetric noise model where the $q$ and $p$ noise quadratures
are both correlated (or both anti-correlated).
\bibitem{fn2} Although the fraction
of the input mean photon number that is due to entanglement $\eta^*\to 0$, its
absolute value $\eta^*\bar{n}$ tends to a constant as $\bar{n}\to\infty$
(except for $x=1$).
\end{thebibliography}
\end{document}